\documentclass[fontsize=11pt,headings=small,headinclude=true,footinclude=true, 
 headings=optiontohead,toc=listof,DIV=22,listof=leveldown, headsepline , letterpaper]{scrartcl} 
\usepackage[utf8]{inputenc}
\usepackage[T1]{fontenc}
\usepackage[ngerman, english]{babel}

\newcommand{\dptitle}{Helmholtz decomposition and potential functions \par for n-dimensional analytic vector fields}
\newcommand{\dptitleclean}{Helmholtz decomposition and potential functions for n-dimensional analytic vector fields}

\newcommand{\dpautoren}{Erhard Glötzl$^1$ \href{https://orcid.org/0000-0002-3092-8243}{\includegraphics[width=1em]{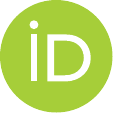}}, Oliver Richters$^{2,3}$ \href{https://orcid.org/0000-0001-8253-4716}{\includegraphics[width=1em]{orcid_logo.pdf}}}
\newcommand{\dpautorenclean}{Erhard Glötzl, Oliver Richters}

\newcommand{\dpaffiliation}{\small 1: Institute of Physical Chemistry, Johannes Kepler University Linz, Austria. \\ 2: Department of Business Administration, Economics and Law, \\ Carl von Ossietzky University, Oldenburg, Germany. \\ 3: Now at: Potsdam Institute for Climate Impact Research, Potsdam, Germany. }

\usepackage{amsmath}
\usepackage{amssymb,amsthm,amsfonts,textcomp}
\usepackage{color, transparent}
\usepackage{dpfloat}
\usepackage{array}
\usepackage{txfonts}
\usepackage{booktabs, multirow}
\usepackage{paralist}
\usepackage{hhline}
\usepackage{graphicx}
\usepackage{hanging}
\usepackage{bm} \usepackage{bbold}
\usepackage[figuresleft]{rotating}
\usepackage{acronym}
\usepackage{microtype}
\usepackage[hang]{footmisc}
\usepackage{changepage} 
\usepackage{appendix}
\usepackage{csquotes}
\usepackage{pdfpages}
\usepackage{xcolor}

\usepackage{pifont} 

\usepackage[backend=biber, natbib=true, style=authoryear-comp, sorting=nyt, url=true, isbn=false, doi=true, eprint=true, maxcitenames=2, citetracker=true, uniquename=false, labeldate=year]{biblatex}

\DeclareSourcemap{
 \maps{
  \map{
   \step[fieldsource=language, fieldset=langid, origfieldval, final]
   \step[fieldset=language, null]
  }
 }
}

\AtEveryBibitem{%
   \clearfield{day}%
   \clearfield{month}
   \clearfield{endday}%
   \clearfield{endmonth}%
}

\usepackage[inner=3.2cm,outer=3.2cm,top=1.2cm,bottom=0.8cm,includeheadfoot , heightrounded]{geometry}

\usepackage{chngcntr}

\makeatletter
\newcommand\arraybslash{\let\\\@arraycr}
\makeatother

\makeatletter
\g@addto@macro\UrlBreaks{\do\*\do\~\do\'\do\"\do\a\do\b\do\c\do\d\do\e\do\f\do\g\do\h\do\i\do\j\do\k\do%
\l\do\m\do\n\do\o\do\p\do\q\do\r\do\s\do\t\do\u\do\v\do\w\do\x\do\y\do\z\do\&\do\1\do\2\do\3\do\4\do\5\do\6\do\7\do\8\do\9\do\0\do\.}
\makeatother

\usepackage{scrlayer-scrpage}

\clearscrheadfoot
\ihead{\pagemark}
\ohead{\scriptsize Glötzl, Richters: Helmholtz decomposition and potential functions for n-dimensional analytic vector fields}
\let\OLDthebibliography\thebibliography
\renewcommand\thebibliography[1]{
 \OLDthebibliography{#1}
 \setlength{\parskip}{0pt}
 \setlength{\itemsep}{0pt plus 0.3ex}
}

\KOMAoption{listof}{leveldown}

\usepackage{ragged2e}
\newcolumntype{L}[1]{>{\raggedright\arraybackslash}p{#1}} 
\newcolumntype{C}[1]{>{\centering\arraybackslash}p{#1}} 
\newcolumntype{R}[1]{>{\raggedleft\arraybackslash}p{#1}} 

\DeclareMathOperator{\grad}{grad}
\DeclareMathOperator{\Div}{div}
\DeclareMathOperator{\curl}{curl}

\DeclareMathOperator{\ROT}{ROT}

\newcommand{\uphi}{u}

\theoremstyle{plain}
\newtheorem{theorem}{Theorem}[section]

\theoremstyle{definition}
\newtheorem{corol}[theorem]{Corollary}
\newtheorem{defin}[theorem]{Definition}
\newtheorem{lemmarev}[theorem]{Lemma}

\newtheorem{example}[theorem]{Example}

\newcommand{\todone}[1]{}
\newcommand{\todo}[1]{}  

\usepackage[hidelinks, breaklinks=true, pdfauthor={\dpautorenclean}, pdftitle={\dptitleclean}, unicode]{hyperref}

\usepackage{tikz}
\usetikzlibrary{shapes.geometric, arrows}
\tikzstyle{startstop} = [rectangle, rounded corners, minimum width=2.8cm, minimum height=1cm,text centered, draw=black, fill=red!30]
\tikzstyle{decision} = [rectangle, rounded corners, minimum width=2.8cm, minimum height=1.2cm, text centered, draw=black, fill=orange!30]
\tikzstyle{process} = [rectangle, minimum width=2.8cm, minimum height=3.2cm, text centered, draw=black, fill=green!30, execute at begin node=]
\tikzstyle{arrow} = [thick,->,>=stealth]

\begin{document}

\selectlanguage{english}

\thispagestyle{scrplain}

\begin{center}
 { \Large { \bfseries \sffamily \dptitle \\ \bigskip  \par \par } \vspace{1em} {\large \dpautoren \par \vspace{1em} \normalsize \dpaffiliation \par \vspace{1em} Version 3 -- February 2023 \par } }
\end{center}
\vspace{1.3em}
\begin{addmargin}{0.05\textwidth}

\textbf{Abstract:}
The Helmholtz decomposition splits a sufficiently smooth vector field into a gradient field and a divergence-free rotation field.
Existing decomposition methods impose constraints on the behavior of vector fields at infinity and require solving convolution integrals over the entire coordinate space.
To allow a Helmholtz decomposition in $\mathbb{R}^n$, we replace the vector potential in $\mathbb{R}^3$ by the rotation potential, an n-dimensional, antisymmetric matrix-valued map describing $n(n-1)/2$ rotations within the coordinate planes.
We provide three methods to derive the Helmholtz decomposition: (1) a numerical method for fields decaying at infinity by using an $n$-dimensional convolution integral, (2) closed-form solutions using line-integrals for several unboundedly growing fields including periodic and exponential functions, multivariate polynomials and their linear combinations, (3) an existence proof for all analytic vector fields.
Examples include the Lorenz and R\"{o}ssler attractor and the competitive Lotka--Volterra equations with $n$ species.

\bigskip


\bigskip

\noindent\begin{minipage}[t]{0.84\linewidth}
\textbf{Licence:} Creative-Commons \href{http://creativecommons.org/licenses/by-nc-nd/4.0/}{CC-BY-NC-ND 4.0}. \end{minipage} 
\begin{minipage}[t]{0.15\linewidth}\vspace{-\ht\strutbox}
\includegraphics[width=\columnwidth]{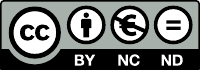}\end{minipage}

\bigskip

\noindent This is the accepted manuscript of the article published in: \newline \emph{Journal of Mathematical Analysis and Applications}, 2023, doi:\href{https://doi.org/10.1016/j.jmaa.2023.127138}{10.1016/j.jmaa.2023.127138}.

\end{addmargin}

\bigskip

\pagestyle{scrheadings}

\clearpage 
\section{Introduction}\label{sec:intro}

The Helmholtz decomposition \citep{stokes_dynamical_1849,  von_helmholtz_uber_1858} splits a sufficiently smooth vector field $\bm{f}$ into an irrotational (curl-free) \emph{gradient field} $\bm{g}$ and a solenoidal (divergence-free) \emph{rotation field} $\bm{r}$.
This `fundamental theorem of vector calculus' is indispensable for many problems in mathematical physics \citep{tran-cong_helmholtzs_1993, kustepeli_helmholtz_2016, dassios_uniqueness_2002}, but also used in animation, computer vision or robotics \citep{bhatia_helmholtz-hodge_2013} and for analyzing the dynamics of complex systems \citep{zhou_quasi-potential_2012, suda_construction_2019}.

The challenge is to derive the \emph{gradient potential} $G$, the \emph{gradient field} $\bm{g} = \grad G$ and the \emph{rotation field} $\bm{r}$ such that:
\begin{align}
\bm{f}(\bm{x}) &= \bm{g}(\bm{x}) + \bm{r}(\bm{x}) = \grad G(\bm{x}) + \bm{r}(\bm{x}), \label{eq_fxex} \\
\Div \bm{r}(\bm{x}) &= 0.
\end{align}
If $\bm{r} = 0$, then $\bm{f}$ is called a \emph{curl-free} vector field, and a line integral yields the gradient potential $G$.\footnote{Note that in physics, $G$ is often defined with an opposite sign.}
In other cases, the Poisson equation $\Delta G(\bm{x}) = \Div \bm{f}(\bm{x})$ has to be solved by numerically computing infinite convolution integrals over $\mathbb{R}^n$ for each point $\bm{x}$.
On bounded domains or for fields decaying sufficiently fast at infinity, a unique solution is guaranteed by appropriate boundary conditions \citep{schwarz_hodge_1995, chorin_mathematical_1990}.
However, for unboundedly growing vector fields, these numerical integrals diverge.
In $\mathbb{R}^3$, the rotation field can also be written as $\curl$ of some \emph{vector potential} $\bm{A}$:
\begin{align}
\bm{r}(\bm{x}) &= \curl \bm{A}(\bm{x}) = \nabla \times \bm{A}(\bm{x}). \label{eq_rot3d}
\end{align}
However, this approach cannot be easily applied to higher dimensions.

\begin{figure}[ht]
\hfil\includegraphics[width=0.8\columnwidth]{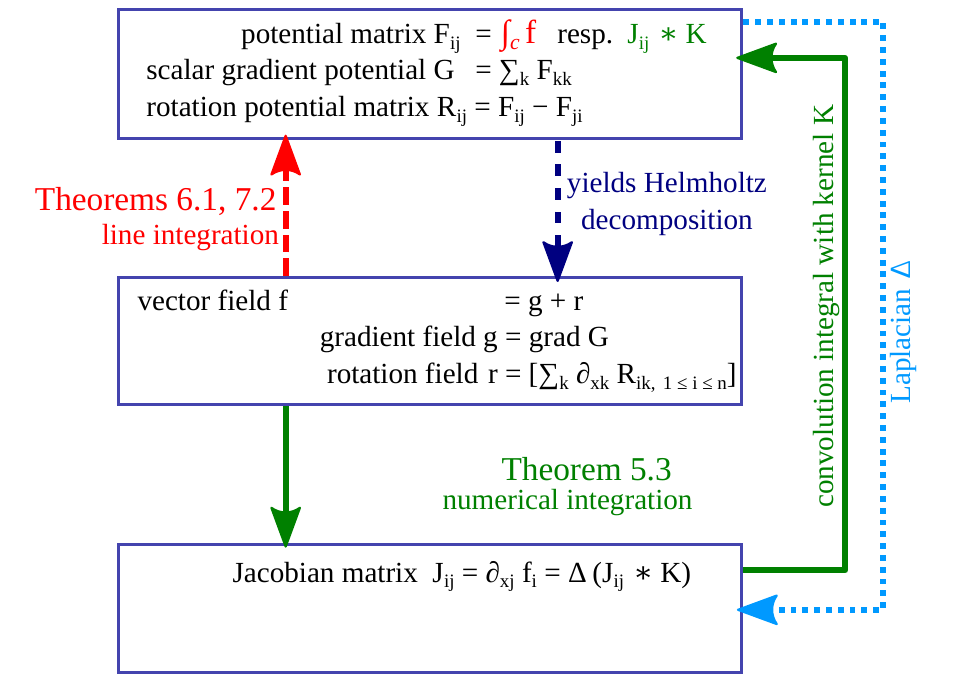}\hfil
\caption{\label{fig_zusammenhang-groessen}The figure depicts how to obtain a Helmholtz decomposition for a vector field $\bm{f}$ by calculating the gradient field and the rotation field from the potential matrix $\mathbf{F} = \llbracket F_{ij} \rrbracket$ (dark blue short-dashed arrow).
The variables and operators are defined in Definitions~\ref{def_helmholtz}--\ref{def_potentialmatrix}. For sufficiently fast decaying fields on unbounded domains, the convolution of the Jacobian $\mathbf{J}$ with a solution $K$ of Laplace's equation numerically yields the potential matrix (Theorem \ref{theorem_hd_decay}, green solid arrows). The Laplacian applied to the potential matrix yields the Jacobian matrix (light blue dotted arrow). The line integration approach (Theorems~\ref{theorem_hd_both} and \ref{theorem_hd_analytic}) allows dropping the condition that the vector field $\bm{f}$ decays sufficiently fast at infinity by directly calculating the potential matrix for analytic vector fields (red long-dashed arrow).}
\end{figure}

This paper overcomes these limitations: We generalize the vector potential $\bm A(\bm x)$ from $\mathbb{R}^3$ to $\mathbb{R}^n$, replacing it by an antisymmetric $n \times n$ matrix-valued map we call \emph{rotation potential} (Section~\ref{sec_theorem}). After comparing this concept to existing approaches (Section~\ref{sec_lowdim}), we provide three methods to derive the Helmholtz decomposition, see Figure \ref{fig_zusammenhang-groessen}:
\begin{itemize}
\item a numerical method for fields decaying at infinity by using an $n$-dimensional convolution integral with a solution of Laplace's equation (Section~\ref{sec_hd_decay}),
\item closed-form solutions for gradient and rotation potentials for several unboundedly growing fields (Section~\ref{sec_hd_intuitive}),
\item an existence proof for all analytic vector fields, locally given by a convergent power series (Section~\ref{sec_hd_analytic}).\todone{Reviewer: -p.2 l.30: "thus" is unnecessary. \par Answer: We deleted it.}
\end{itemize}
Section~\ref{sec_conclusion} discusses our results and the Appendix presents five examples.\todone{Reviewer: -p.2 l.32: "the appendix" -> "Appendix" \par Answer: thanks.}

\section{Notation}
\label{sec_notation}

Let $f$ be a scalar function.
We denote the partial derivative as $\partial_{x_j} f = \tfrac{\partial f}{\partial_{x_j}}$, the $p$-th partial derivative as $\partial_{x_j}^p f$, the Laplace operator as $\Delta$ and the Laplacian to the power of $p$ as $\Delta^p f \coloneqq \Delta (\Delta^{p-1} f)$ for $p \in \mathbb{N}$.

We define the incomplete Laplacian and its $p$-th power as:\todone{Reviewer: -p.3 l.15: "p-the" -> "p-th" \par Answer: thanks}
\begin{align}
\Delta_{\setminus k} f \coloneqq \sum_{i \neq k} \partial_{x_i}^2 f \quad \text{ and } \quad \Delta_{\setminus k}^p f \coloneqq \Delta_{\setminus k} (\Delta_{\setminus k}^{p-1} f).
\end{align}

We denote the antiderivative of $f$ with respect to $x_j$ as:
\begin{align}
\mathcal{A}_{x_j} f(\bm{x}) \coloneqq& \int_0^{x_j} f(\bm{\xi}) d\xi_j,
\end{align}
and the $p$-th antiderivative with respect to $x_j$ as the Riemann--Liouville integral\footnote{Also called Cauchy formula for repeated integration \citep{riesz_integrale_1949, cauchy_trente-cinquieme_1823}.}:
\begin{align}
\mathcal{A}_{x_j}^p f(\bm{x}) \coloneqq&\, \mathcal{A}_{x_j} \mathcal{A}_{x_j}^{p-1} f(\bm{x}) = \frac{1}{(p-1)!} \int_0^{x_j} (x_j - \xi_j)^{p-1} f(\bm{\xi}) d\xi_j, 
\end{align}
such that $\partial_{x_j} \mathcal{A}_{x_j} f = \partial_{x_j}^p \mathcal{A}_{x_j}^p f = f$.
By convention, $\partial_{x_j}^0 f = \Delta^0 f = \Delta_{\setminus k}^0 f = \mathcal{A}_{x_j}^0 f = f$.

We denote an $n$-dimensional vector field $\bm{f} = [f_k; \,_{1 \leq k \leq n}]$ with bold, italic font and use single indices for its components $f_k$.
If a function $f(\bm x)$ does not depend on a coordinate $x_m$, we denote it as $f(\bm x_{i \neq m})$.
We denote an $n \times n$ matrix-valued map $\mathbf{L} = \llbracket L_{ij} \rrbracket = \llbracket L_{ij}; \,_{1 \leq i \leq n, 1 \leq j \leq n} \rrbracket$ with bold, non-italic font.
For vector fields and matrices, the definitions for derivatives and integrals above are applied component-wise.

For $h \in \mathbb{R}$, $\lceil h \rceil$ is the least integer greater than or equal to $h$ (ceiling function).

\section{Helmholtz decomposition using n-dimensional potential matrices}
\label{sec_theorem}

First, we define the terms of the paper, the relations of which are illustrated in Figure~\ref{fig_zusammenhang-groessen}.

\begin{defin}[Helmholtz decomposition, gradient field $\bm g$ and rotation field $\bm r$] \label{def_helmholtz}
For a vector field $\bm f \in C^1(\mathbb{R}^n, \mathbb{R}^n)$, a \emph{Helmholtz decomposition} is a pair of vector fields $\bm g \in C^1(\mathbb{R}^n, \mathbb{R}^n)$ and $\bm{r} \in C^1(\mathbb{R}^n, \mathbb{R}^n)$ such that
$\bm{f} = \bm{g} + \bm{r}$, where $\bm{g} = \grad G$ for a function $G \in C^2(\mathbb{R}^{n}, \mathbb{R})$ and $\Div \bm{r} = 0$. The vector field $\bm{g}$ is called a \emph{gradient field} and $\bm{r}$ is called a \emph{rotation field}.
\end{defin}

\begin{defin}[Rotation operators $\ROT$ and $\overline{\ROT}$] \label{def_ROT}
We define two \emph{rotation operators}, $\overline{\ROT}$ mapping a vector field $\bm{f} \in C^1(\mathbb{R}^n, \mathbb{R}^n)$ to a matrix-valued map, and $\ROT$ mapping a matrix-valued map $\mathbf{R} \in C^1(\mathbb{R}^n, \mathbb{R}^{n^2})$ with elements $R_{ij}$ to a vector field:
\begin{align}
  \overline{\ROT} \bm{f}(\bm{x}) &\coloneqq \left\llbracket \partial_{x_j} f_i(\bm{x}) - \partial_{x_i} f_j(\bm{x}); \,_{1 \leq i \leq n, 1 \leq j \leq n} \right\rrbracket. \\
 \ROT \mathbf{R}(\bm{x}) &\coloneqq \left[ \sum\nolimits_k \partial_{x_k} R_{ik}(\bm{x});\,_{1 \leq i \leq n} \right].
\end{align}
The latter is equivalent to the divergence performed with respect to the rows of $\mathbf{R}$.
\end{defin}

\begin{defin}[Potential matrix $\mathbf{F}$, gradient potential $G$, rotation potential $\mathbf{R}$] \label{def_potentialmatrix}

For each vector field $\bm{f} \in C^1(\mathbb{R}^n, \mathbb{R}^n)$, a matrix-valued map $\mathbf{F} = \llbracket F_{ij} \rrbracket \in C^2(\mathbb{R}^n, \mathbb{R}^{n^2})$ is called a \emph{potential matrix} of $\bm{f}$, a scalar function $G \in C^2(\mathbb{R}^{n}, \mathbb{R})$ is called a \emph{gradient potential} and a matrix-valued map $\mathbf{R}  = \llbracket R_{ij} \rrbracket \in C^2(\mathbb{R}^n, \mathbb{R}^{n^2})$ is called a \emph{rotation potential} of $\bm{f}$ if three conditions are met:
Firstly, $G$ is the sum of the diagonal elements of $\mathbf{F}$ and, secondly, $\mathbf{R}$ is equal to $\mathbf{F} - \mathbf{F}^\top$:
\begin{align}
G(\bm{x}) &= \sum_k F_{kk}(\bm{x}), \label{eq_def_G} \\
R_{ik}(\bm{x}) &= F_{ik}(\bm{x}) - F_{ki}(\bm{x}). \label{eq_def_R}
\end{align}
Thirdly, the gradient field $\bm{g}$ and the rotation field $\bm{r}$ as defined below yield a Helmholtz decomposition of $\bm{f}$ such that $\bm{f}(\bm{x}) = \bm{g}(\bm{x}) + \bm{r}(\bm{x})$:
\begin{align}
\bm{g}(\bm{x}) \coloneqq& \grad G(\bm{x}) = \left[ \partial_{x_i} G(\bm{x}),\,_{1 \leq i \leq n}\right] = \left[ \partial_{x_i} \sum\nolimits_k F_{kk}(\bm{x}),\,_{1 \leq i \leq n}\right], \label{eq_def_g} \\
\bm{r}(\bm{x}) \coloneqq& \ROT \mathbf{R}(\bm{x}) = \left[ \sum\nolimits_k \partial_{x_k} R_{ik}(\bm{x});\,_{1 \leq i \leq n} \right] = \left[ \sum\nolimits_k \partial_{x_k} \Big( F_{ik}(\bm{x}) - F_{ki}(\bm{x}) \Big) ;\,_{1 \leq i \leq n} \right]. \label{eq_def_r}
\end{align}
\end{defin}

Note that the rotation potential $\mathbf{R}$ as defined in Eq.~\eqref{eq_def_R} is antisymmetric and therefore contains $\binom{n}{2}$ independent entries.
The following Lemma~\ref{lemma_hd} shows that, for any $C^2$ matrix-valued map $\mathbf{F}$, a vector field $\bm{f}$ and its Helmholtz decomposition can be derived. In Sections \ref{sec_hd_decay}--\ref{sec_hd_analytic}, we develop methods for the inverse problem to calculate such a potential matrix $\mathbf{F}$ if the vector field $\bm{f}$ is given.
We present methods to calculate $\mathbf{F}$ numerically by solving convolution integrals, and to derive the potential matrix using line integration, see Figure \ref{fig_zusammenhang-groessen}.

\begin{lemmarev}  For every matrix-valued map $\mathbf{F} \in C^2(\mathbb{R}^n, \mathbb{R}^{n^2})$, $\bm{g}$ as defined by Eq.~\eqref{eq_def_g} is a gradient field with $\overline{\ROT} g(\bm x) = 0$, and the rotation field $\bm{r}$ as defined by Eq.~\eqref{eq_def_r} is divergence-free, satisfying $\Div \bm r(\bm x) = 0$.
If $\bm{f} = \bm{g} + \bm{r}$, the matrix-valued map $\mathbf{F}$ is a \emph{potential matrix} of $\bm{f}$. \label{lemma_hd}
\end{lemmarev}
\begin{proof}
$\bm g$ is a gradient field by definition and the statements above can be proven by simple calculation:
\begin{align}
\begin{split}
\overline{\ROT} \bm g(\bm x) &= \left\llbracket \partial_{x_j} g_i(\bm{x}) - \partial_{x_i} g_j(\bm{x}); \,_{1 \leq i \leq n, 1 \leq j \leq n} \right\rrbracket \\
&= \left\llbracket \partial_{x_j} \partial_{x_i} G(\bm{x}) - \partial_{x_i} \partial_{x_j} G(\bm{x}); \,_{1 \leq i \leq n, 1 \leq j \leq n} \right\rrbracket = 0,
\end{split} \\
\begin{split} \Div \bm{r}(\bm{x}) &= \sum\nolimits_i \partial_{x_i} r_i(\bm{x}) = \sum\nolimits_i \partial_{x_i} \sum\nolimits_k \partial_{x_k} R_{ik}(\bm{x}) \\
&= \sum_{i,k} \partial_{x_i} \partial_{x_k} (F_{ik}(\bm{x}) - F_{ki}(\bm{x})) = 0, \end{split} \label{eq_divr}
\end{align}
because the partial derivatives can be exchanged. Therefore, the matrix $\mathbf{F}$ is a potential matrix of the vector field $\bm{f}$ given by $\bm{f} = \bm{g} + \bm{r}$.
\end{proof}

\begin{lemmarev} \label{lemma_linear} 
Let $\bm{f^{(1)}}, \bm{f^{(2)}} \in C^1(\mathbb{R}^n, \mathbb{R}^n)$ be two vector fields with corresponding potential matrices $\mathbf{F^{(1)}}$ and $\mathbf{F^{(2)}}$, and $h \in \mathbb{R}$.
Then, $\mathbf{F} = h \mathbf{F^{(1)}} + \mathbf{F^{(2)}}$ is a potential matrix for the vector field $\bm{f} = h \bm{f^{(1)}} + \bm{f^{(2)}}$.
\end{lemmarev}
\begin{proof}
The proof follows directly by linearity of all the operators in Definitions~\ref{def_helmholtz}--\ref{def_potentialmatrix}.
\end{proof}

\section{Comparison to existing approaches} \label{sec_lowdim}


Let us now compare the $n$-dimensional rotation potential to existing approaches \citep[see also][]{glotzl_helmholtz_2021}. 
If the vector potential $\bm{A}$ in Eq.~\eqref{eq_rot3d} is known, its entries can be used as the three independent matrix elements of the rotation potential such that the rotation field is the same:
\begin{align}
\mathbf{R} &= \begin{bmatrix} 0 & +A_3 & -A_2 \\ -A_3 & 0 & +A_1 \\ +A_2 & -A_1 & 0 \end{bmatrix}, \qquad \bm{A} = \begin{bmatrix} A_1 \\ A_2 \\ A_3 \end{bmatrix} = \begin{bmatrix} R_{23} \\ R_{31} \\ R_{12} \end{bmatrix} = \begin{bmatrix} F_{23} - F_{32} \\ F_{31} - F_{13} \\ F_{12} - F_{21} \end{bmatrix}, \\
\bm{r}(\bm{x}) &= \nabla \times \bm{A}(\bm{x}) = \begin{bmatrix} \partial_{x_2} A_3 - \partial_{x_3} A_2 \\ \partial_{x_3} A_1 - \partial_{x_1} A_3 \\ \partial_{x_1} A_2 - \partial_{x_2} A_1 \end{bmatrix} \equiv \begin{bmatrix} \partial_{x_2} R_{12} + \partial_{x_3} R_{13} \\  \partial_{x_1} R_{21} + \partial_{x_3} R_{23} \\  \partial_{x_1} R_{31} + \partial_{x_2} R_{32} \end{bmatrix} = \ROT \mathbf{R}(\bm{x}),
\end{align}
using $R_{ij} = -R_{ji}$.
Writing the rotation potential as a vector potential is possible only in $\mathbb{R}^3$, because $\binom{n}{2} = n$ if and only if $n = 3$.

Existing extensions of the rotation potential to $\mathbb{R}^n$ use a tensor field $T$ with dimension $n$ and rank $n - 2$ to get $\bm{r}(\bm{x}) = \curl T(\bm{x})$.\footnote{The $\curl$ of a tensor $T_\alpha$ of rank $n-2$, described by a multi-index $\alpha$ of dimension $n-2$, can be defined as: $\bm f(\bm x) = [f_1, f_2, \dots, f_n] = \curl T_\alpha$ with $f_k(\bm x) = \sum_{i = 1}^n \partial_{x_i} \epsilon_{ki\alpha} T_\alpha(\bm x)$ using the Levi-Civita epsilon \citep[Eq.~1.5 with $m=n-2$]{georgievskii_second_2020}. The definition used in \emph{Mathematica} \citep{wolfram_research_curl_2012} uses an additional factor $1/(n-2)!$ on the right hand side.}
This tensor field $T$ therefore has $n^{n-2}$ components, a complexity that increases with $n$ faster than exponentially.
Our matrix approach that requires only $n^2$ entries substantially reduces complexity and improves tractability for $n>4$.

In two dimensions, to get $\bm{r}(\bm{x}) =  \curl T(\bm x)$, $T$ needs to be a scalar function:\footnote{See the definition in \citet[p.~809]{georgievskii_second_2020}, Eq.~(1.5) in $\mathbb{R}^2$ with rank $m = 0$.}
\begin{align}
\bm{r}(\bm{x}) = \curl T(\bm x) = \left[- \partial_{x_2} T(\bm x), \partial_{x_1} T(\bm x) \right] = \left[ \delta_{2k} \partial_{x_1} T(\bm x) - \delta_{1k} \partial_{x_2} T(\bm x);\,_{1 \leq k \leq 2} \right]. \label{eq_2dcurl_scalar}
\end{align}
In our approach, the function $T$ appears in all non-diagonal entries of the rotation potential $\mathbf{R}(\bm{x}) = \begin{bmatrix} 0 & +T(\bm x) \\ -T(\bm x) & 0 \end{bmatrix}$.
The matrix component $R_{12}(\bm x) = T(\bm x)$ describes the rotation within the $x_1$-$x_2$-plane in $\mathbb{R}^2$ and, similarly, in $\mathbb{R}^n$ each matrix component $R_{ij}$ describes the rotation within the $x_i$-$x_j$-plane:
If only one single matrix element $R_{ij} = -R_{ji}$ is non-zero, the rotation field $\bm{r}$ is given by:
\begin{align}
\bm{r}(\bm{x}) \coloneqq& \left[ \delta_{ik} \partial_{x_j} R_{ij}(\bm x) + \delta_{jk} \partial_{x_i} R_{ji}(\bm x);\,_{1 \leq k \leq n} \right],
\end{align}
with the Kronecker delta $\delta_{ik} = 1$ if $i = k$ and $0$ otherwise.
This vector describes the rotation within the $x_i$-$x_j$-plane, as only components at the positions $i$ and $j$ can be non-zero, taking the values $\partial_{x_j} R_{ij}$ and $\partial_{x_i} R_{ji}$.

For a general rotation, each of the $\binom{n}{2}$ independent matrix entries of the rotation potential $\mathbf{R}$ describes one of the rotations within the coordinate planes, and the rotation field is the sum of the respective vectors:
\begin{align}
\sum_{i > j} \left[ \delta_{ik} \partial_{x_j} R_{ij} +  \delta_{jk} \partial_{x_i} R_{ji};\,_{1 \leq k \leq n} \right] = \left[ \sum_{m=1}^n \partial_{x_m} R_{km};\,_{1 \leq k \leq n} \right], \label{eq_def_ROT}
\end{align}
which is equivalent to the definition of $\bm{r}(\bm{x})$ as $\ROT \mathbf{R}(\bm x)$ in Eq.~\eqref{eq_def_r}.

There exists a well-known identity relating the Laplacian $\Delta$, $\Div$, $\grad$ and $\curl$ in $\mathbb{R}^3$ \citep[p.~522]{de_la_calle_ysern_constructive_2019}:
\begin{align}
\grad \Div \bm{f}(\bm{x}) - \curl \curl \bm{f}(\bm{x}) = \Delta \bm{f}(\bm{x}).
\end{align}
We show that a similar identity exists for $\ROT$ and $\overline{\ROT}$ in $\mathbb{R}^n$.

\begin{lemmarev} \label{lemma_laplace_identity} 
For any twice continuously differentiable vector field $\bm{f} = \left[ f_k(\bm{x}); \,_{1 \leq k \leq n}\right]$, the following identity holds:
\begin{align}
\grad \Div \bm{f}(\bm{x}) + \ROT \overline{\ROT} \bm{f}(\bm{x}) = \Delta \bm{f}(\bm{x}).
\end{align}
\end{lemmarev}
\begin{proof} Using the definition of $\ROT$ and $\overline{\ROT}$ results in the left hand side being:
\begin{align}
\begin{aligned}
\text{L.H.S.} &= \grad \Div \bm{f}(\bm{x}) + \ROT \left\llbracket \left( \partial_{x_j} f_i(\bm{x}) -  \partial_{x_i} f_j(\bm{x}) \right); \,_{1 \leq i \leq n, 1 \leq j \leq n} \right\rrbracket \\
&= \left[ \partial_{x_k} \sum_{m=1}^n \partial_{x_m} f_m(\bm{x}) + \sum_{m=1}^n \partial_{x_m} \Big( \partial_{x_m} f_k(\bm{x}) - \partial_{x_k} f_m(\bm{x}) \Big); \,_{1 \leq k \leq n} \right] \\
&= \left[\sum_{m=1}^n \partial_{x_m}^2 f_k(\bm{x}); \,_{1 \leq k \leq n} \right] = \left[ \Delta f_k(\bm{x}); \,_{1 \leq k \leq n} \right] = \Delta \bm{f}(\bm{x}). \label{eq_laplace_identity1}
\end{aligned}
\end{align}
The terms cancel each other because of the symmetry of second derivatives (Schwarz's theorem) and interchangeability of sums and derivatives.
\end{proof}
In the following three sections, we show how to derive potential matrices for different vector fields.

\section{Helmholtz theorem for decaying fields}
\label{sec_hd_decay}

In $\mathbb{R}^3$, for vector fields $\bm{f}$ that decay sufficiently fast at infinity, $G$ and $\bm A$ can be derived using Kernel integrals.
After defining the Kernel integral and proving a Lemma that allows to interchange it with a partial derivative, we show in Theorem~\ref{theorem_hd_decay} how Kernel integrals can be used to derive the potentials used for our approach in $\mathbb{R}^n$.

\begin{defin}[Kernel integral] \label{def_kernel}
Let $q$ be an $n$-dimensional $C^1$ vector field that decays faster than $\left|x\right|^{-d}$ for $\left|x\right| \to \infty$ with $d > 2$.
Let the integral kernel $K$ be given by the fundamental solution of Laplace's equation:
\begin{align}
K(\bm{x}, \bm{\xi}) &= \begin{cases}
         \frac{1}{2\pi}  \log{ | \bm{x} - \bm{\xi} | } &  n=2,  \\
         \frac{1}{n(2-n)V_n} | \bm{x} - \bm{\xi} | ^{2-n} &  \text{for $n = 1$ or $n \geq 3$},
      \end{cases} \label{eq_kernel}
\end{align}
with $V_n = \pi^\frac{n}{2} / \Gamma\big(\tfrac{n}{2}+1\big)$ being the volume of a unit $n$-ball and the gamma function $\Gamma$.
Then, $Q$, the \emph{Kernel integral} of $q$, is defined as the convolution of $q$ with $K$, given by:
\begin{align}
Q(\bm x) = \int_{\mathbb{R}^n} q(\bm{\xi}) K(\bm{x}, \bm{\xi}) d\bm{\xi}^n. \label{eq_Kernel}
\end{align}
\end{defin}
From the theory of the Poisson equation, it is well-known that Eq.~\eqref{eq_Kernel} implies $\Delta Q(\bm x) = q(\bm x)$.

\begin{lemmarev} \label{lemma_kernel_partial}
Let $q$ be a $C^1$ vector field that decays faster than $\left|x\right|^{-d}$ for $\left|x\right| \to \infty$ with $d > 2$. 
Then, any partial derivative $\partial_{x_k}$ of the Kernel integral of $q$ is identical to the Kernel integral of $\partial_{x_k} q$:
\begin{align}
\partial_{x_k} \int_{\mathbb{R}^n} q(\bm{\xi}) K(\bm{x}, \bm{\xi}) \,d\bm{\xi}^n = \int_{\mathbb{R}^n} \frac{\partial q(\bm{\xi})}{\partial \xi_k} K(\bm{x}, \bm{\xi}) \,d\bm{\xi}^n.
\end{align}
\begin{proof}
\begin{align}
\begin{split} & \partial_{x_k} \int_{\mathbb{R}^n} q(\bm{\xi}) K(\bm{x}, \bm{\xi}) \,d\bm{\xi}^n = \int_{\mathbb{R}^n} q(\bm{\xi}) \frac{\partial K(\bm{x}, \bm{\xi})}{\partial_{x_k}} \,d\bm{\xi}^n = \int_{\mathbb{R}^n} q(\bm{\xi}) \frac{ - \partial K(\bm{x}, \bm{\xi})}{\partial_{\xi_k}} \,d\bm{\xi}^n \end{split} \label{eq_lemma_kernel_partial_2} \\
\intertext{applying integration by parts in the $k$-component}
=& - \int_{\mathbb{R}^{n-1}} \left( q(\bm{\xi}) K(\bm{x}, \bm{\xi}) \Big|_{\xi_k = -\infty}^{\infty}  - \int_{\xi_k=-\infty}^{\infty} \frac{\partial q(\bm{\xi})}{\partial_{\xi_k}} K(\bm{x}, \bm{\xi})  \,d\xi_k \right) \,d[\xi_i; i \neq k]  \label{eq_lemma_kernel_partial_3} \\
\intertext{and, as in the first summand, we have $q(\bm{\xi}) K(\bm{x}, \bm{\xi}) \to 0$ for $\xi_k \to \pm \infty$}
=& \int_{\mathbb{R}^n} \frac{\partial q(\bm{\xi})}{\partial_{\xi_k}} K(\bm{x}, \bm{\xi}) \,d\bm{\xi}^n.
\end{align}
For $n=1$, the outer integral over $\mathbb{R}^0$ in Eq.~\eqref{eq_lemma_kernel_partial_3} is omitted.
\end{proof}
\end{lemmarev}

In $\mathbb{R}^3$, if $\bm f$ satisfies the conditions of Def.~\ref{def_kernel}, $G$ and $\bm A$ can be derived using Kernel integrals \citep{blumenthal_uber_1905, petrascheck_helmholtz_2015}:
\begin{align}
G(\bm x) & = \int_{\mathbb{R}^3} \Div \bm f(\bm \xi) \cdot K(\bm x, \bm \xi) \, \mathrm{d}\bm \xi^3, \\
\bm A(\bm x) & = \int_{\mathbb{R}^3} \curl \bm f(\bm \xi) \cdot K(\bm x, \bm \xi) \, \mathrm{d}\bm \xi^3,
\end{align}
satisfying the two Poisson equations $\Delta G(\bm{x}) = \Div \bm{f}(\bm{x})$ and $\Delta \bm{A}(\bm{x}) = \curl \bm{f}(\bm{x}) = \nabla \times \bm{f}(\bm{x})$.
A similar approach for $\mathbb{R}^n$ will be presented in the following.

\begin{theorem}[Helmholtz theorem for decaying fields] \label{theorem_hd_decay} 
Let $\bm{f} \in C^2(\mathbb{R}^n, \mathbb{R}^n)$ be a vector field that decays faster than $|\bm{x}|^{-d}$ for $|\bm{x}| \to \infty$ with $d > 2$.
Then, the matrix $\mathbf{F} = \llbracket F_{ij} \rrbracket$ defined as the Kernel integral of the Jacobian matrix of $\bm{f}$ is a potential matrix of $\bm{f}$:
\begin{align}
F_{ij}(\bm{x}) &= \int_{\mathbb{R}^n} \big( \partial_{\xi_j} f_i(\bm{\xi}) \big) K(\bm{x}, \bm{\xi}) d\bm{\xi}^n. \label{eq_F_decaying}
\end{align}
The corresponding gradient potential $G$ and rotation potential $\mathbf{R}$ satisfy $\Delta G(\bm x) = \Div \bm f(\bm x)$ and $\Delta \mathbf{R}(\bm{x}) = \overline{\ROT} \bm{f}(\bm{x})$.
\end{theorem}

\begin{proof}
The gradient potential $G$ and the rotation potential $\mathbf{R} = \llbracket R_{ij} \rrbracket$ are given by:
\begin{align}
\begin{split}
  G(\bm{x}) &= \sum_i F_{ii}(\bm x) = \sum_i \int_{\mathbb{R}^n} \Big(\partial_{\xi_i} f_i(\bm{\xi}) \Big) K(\bm{x}, \bm{\xi}) d\bm{\xi}^n \\
    &= \int_{\mathbb{R}^n} \Big(\sum_i  \partial_{\xi_i} f_i(\bm{\xi}) \Big) K(\bm{x}, \bm{\xi}) d\bm{\xi}^n = \int_{\mathbb{R}^n} \Big(\Div \bm{f}(\bm \xi) \Big) K(\bm{x}, \bm{\xi}) d\bm{\xi}^n.
\end{split} \\
\begin{split}
  R_{ij}(\bm{x}) &= F_{ij}(\bm x) - F_{ji}(\bm x)
    = \int_{\mathbb{R}^n} \Big( \partial_{\xi_j} f_i(\bm{\xi}) - \partial_{\xi_i} f_j(\bm{\xi}) \Big) K(\bm{x}, \bm{\xi}) d\bm{\xi}^n, \\
\Leftrightarrow  \mathbf{R}(\bm x) &= \int_{\mathbb{R}^n} \Big( \overline{\ROT} \bm{f}(\bm{\xi}) \Big) K(\bm{x}, \bm{\xi}) d\bm{\xi}^n.
\end{split}
\end{align}
The construction with the Kernel integral implies that $\Delta G(\bm x) = \Div \bm f(\bm x)$ and $\Delta \mathbf{R}(\bm{x}) = \overline{\ROT} \bm{f}(\bm{x})$ which is equivalent to $\Delta R_{ij}(\bm{x}) = \partial_{x_j} f_i(\bm{x}) - \partial_{x_i} f_j(\bm{x})$.

Lemma~\ref{lemma_hd} ensures that $\bm{g}$ is a gradient field and $\bm{r}$ is divergence-free.
It remains to be shown that $\bm{f} = \bm{g} + \bm{r}$ in each component:
\begin{align}
 g_i(\bm{x}) + r_i(\bm{x}) &= \partial_{x_i} \sum_{k=1}^n F_{kk}(\bm{x}) + \sum_{k=1}^n \partial_{x_k} (F_{ik}(\bm{x}) - F_{ki}(\bm{x})), \label{eq_g_r_decay}
\end{align}
which can be simplified to one single integral using Lemma \ref{lemma_kernel_partial}:
\begin{align}
 \begin{split} &= \int_{\mathbb{R}^n} \left(  \partial_{\xi_i} \sum_{k=1}^n \partial_{\xi_k} f_k(\bm{\xi}) + \sum_{k=1}^n \partial_{\xi_k} \Big( \partial_{\xi_k} f_i(\bm{\xi}) - \partial_{\xi_i} f_k(\bm{\xi}) \Big)  \right) K(\bm{x}, \bm{\xi}) d\bm{\xi}^n \\
  &= \int_{\mathbb{R}^n} \Big(  \Delta f_i(\bm{\xi}) \Big) K(\bm{x}, \bm{\xi}) d\bm{\xi}^n = f_i(\bm{x}), \end{split} 
\end{align}
using $\int_{\mathbb{R}^n} \left( \Delta q(\bm \xi) \right) K(\bm{x}, \bm{\xi}) d\bm{\xi}^n = q(\bm{x})$.
\end{proof}

The sufficiently fast decay of $\bm{f}$ at infinity is required to ensure that the integral in Eq.~\eqref{eq_F_decaying} converges and the potentials are well-defined.
Previous attempts to increase the applicability used more complicated kernel functions such as $K(\bm{x}, \bm{\xi}) = \frac{1}{2\pi} \left( \log{ | \bm{x} - \bm{\xi} | } - \log{|\bm{\xi} | } \right)$ for $n=2$, and $\frac{1}{n(2-n)V_n} \left( | \bm{x} - \bm{\xi} | - |\bm{\xi}| \right) ^{2-n}$ otherwise.
Then, the elements of the Jacobian must decay faster than  $|\bm{x}|^{-d}$ with $d > 0$ only, instead of $d > 2$.
Such a numerical integration method can also be applied to functions that grow slower than a polynomial $|\bm{x}|^{-d}$ with $d < 0$ by using much more complicated kernel functions \citep{blumenthal_uber_1905, petrascheck_helmholtz_2015, tran-cong_helmholtzs_1993}.

In the following, we do not improve the kernel, but rather describe a completely different methodology to derive solutions for analytic functions, without restrictions concerning their behavior at infinity.

\section{Closed-form solutions for non-decaying vector fields}
\label{sec_hd_intuitive}

In this section, we will show how to obtain a closed-form solution of the potential matrix for many cases of unboundedly growing fields.
In Section~\ref{sec_hd_analytic}, we show that such a decomposition exists for all analytic vector fields.
To give you an introduction to our approach, let us discuss a vector field $\bm{f}(\bm{x}) = [ f_1(\bm{x}), 0, \dots, 0 ]$, which allows to set all matrix components $F_{ij}$ with $i \neq 1$ to zero.
Therefore, the gradient potential is given by $G(\bm{x}) = F_{11}(\bm{x})$ and the rotation potential $\mathbf{R}$ is zero except in the first row and column.
To make sure that $g_{i \neq 1}(\bm x) + r_{i \neq 1}(\bm x) = 0$, we try to compensate terms created by calculating a gradient component such as $g_2(\bm x) = \partial_{x_2} G(\bm x)$ by adjusting $R_{12}(\bm x)$ of the rotation potential. Vice versa, we try to compensate terms created by $\ROT \mathbf{R}$ by adjusting $G$.
This can be achieved by choosing $F_{11}(\bm{x})$ such that $\partial_{x_1} F_{11}(\bm{x}) = f_1(\bm{x})$, and $F_{1k}(\bm{x})$ for $k \neq 1$ such that $\partial_{x_1} F_{1k}(\bm{x}) = \partial_{x_k} F_{11}(\bm{x})$.
Then, 
\begin{align}
\bm{g}(\bm{x}) &= \grad G(\bm{x}) = \Big[ \partial_{x_1} F_{11}(\bm{x}), \partial_{x_2} F_{11}(\bm{x}), \dots, \partial_{x_n} F_{11}(\bm{x}) \Big], \\
\bm{r}(\bm{x}) &= \Big[ \sum\nolimits_{k \neq 1} \partial_{x_k} F_{1k}(\bm{x}), - \partial_{x_1} F_{12}(\bm{x}), \dots, - \partial_{x_1} F_{1n}(\bm{x}) \Big].
\end{align}
The sum $\bm{g} + \bm{r}$ is zero except in the first component, although $\bm{g}$ and $\bm{r}$ individually will usually have multiple non-zero components.
If a function $W(\bm x)$ exists such that $\partial_{x_1}^2 W(\bm x) = f_1(\bm x)$, which can often be obtained by line integration $W(\bm x) = \mathcal{A}_{x_1}^2 f_1(\bm x)$, an appropriate choice is:
\begin{align}
F_{1j}(\bm x) &= \partial_{x_j} W(\bm x). \label{eq_example_FU}
\end{align}
Then, the first component of $\bm g(\bm x) + \bm r(\bm x)$ is given by:
\begin{align}
g_1(\bm{x}) + r_1(\bm{x}) &= f_1(\bm{x}) + \sum_{k \neq 1} \partial_{x_k}^2 W(\bm x) = f_1(\bm{x}) + \Delta_{\setminus 1} W(\bm x), \label{eq_multiple}
\end{align}
using the notation with the incomplete Laplacian.
If this term is a multiple $\uphi \neq 0$ of $f_1(\bm{x})$, then a potential matrix of $\bm f$ is obtained by dividing $F_{ij}$ by $\uphi$ in Eq.~\eqref{eq_example_FU}.\footnote{Note that in this process, we have basically integrated twice with respect to $x_1$ to get $W$, while taking the second derivative with respect to all other coordinates in Eq.~\eqref{eq_multiple}. Integrating one part while differentiating the other is a concept known from repeated integration by parts, where the calculation can be finalized as well only if the result yields a multiple of the original function \citep[p.~173]{polyanin_concise_2010}.}

For example, the exponentially growing vector field $\bm{f} \in C^\infty(\mathbb{R}^2, \mathbb{R}^2)$
\begin{align}
 \bm{f}(x_1, x_2) = \Big[ f_1(\bm{x}), 0 \Big] = \Big[ e^{x_1} e^{x_2}, 0 \Big], \label{eq_example_exp}
\end{align}
with $W(\bm x) = e^{x_1} e^{x_2}$ and $F_{1j}$ defined by Eq.~\eqref{eq_example_FU} yields:
\begin{align}
\mathbf{F}(\bm x) &= \begin{pmatrix} \partial_{x_1} W(\bm x) & \partial_{x_2} W(\bm x) \\ 0 & 0 \end{pmatrix} = \begin{pmatrix} e^{x_1} e^{x_2} & e^{x_1} e^{x_2} \\ 0 & 0 \end{pmatrix}, \\
\bm{g}(\bm{x}) &= \big[ + e^{x_1} e^{x_2}, + e^{x_1} e^{x_2} \big],  \\
\bm{r}(\bm{x}) &=  \big[ + e^{x_1} e^{x_2}, - e^{x_1} e^{x_2} \big].
\end{align}
Because $\bm{g} + \bm{r}$ equals $2 \bm{f}$, dividing $\mathbf{F}$ by $\uphi = 2$ yields a potential matrix of $\bm f$.

If the result in Eq.~\eqref{eq_multiple} was not a multiple $\uphi$ of $f_1(\bm{x})$, one can start over again, try to determine a potential matrix for $\left[ \Delta_{\setminus 1} W(\bm x), 0, \dots, 0 \right]$, and substract the result from $\mathbf{F}$ to get a potential matrix of $\bm{f}$.
In Theorem~\ref{theorem_hd_both}, we provide closed-form solutions for the case when repeating this process $\lambda$-times yields a multiple of $f_1(\bm{x})$,
while further generalizing the procedure for antiderivatives with respect to other coordinates, such that $\partial_{x_m}^{2\lambda} W(\bm x) = f_k(\bm x)$, and to linear combinations which includes fields with multiple non-zero coordinates.
Three corollaries provide simplified formulas for special cases.

\begin{theorem}[Helmholtz theorem for unbounded fields] \label{theorem_hd_both}
Let the vector field $\bm{f} \in C^2(\mathbb{R}^n, \mathbb{R}^n)$ be non-zero in only one coordinate $k$, thus $\bm f(\bm x) = [0, \dots, 0, f_k(\bm{x}), 0, \dots, 0]$, and let there exist a natural number $\lambda$, a non-zero real number $\uphi$ and a $C^{2+2\lambda}$ function $W(\bm{x})$ with a $W$-coordinate $x_m$, such that:
\begin{align}
\partial_{x_m}^{2 \lambda} W(\bm{x}) &= f_k(\bm{x}), \label{eq_bedingung_W1} \\
(-1)^{\lambda} \Delta_{\setminus m}^{\lambda} W(\bm{x}) &= (1 - \uphi) f_k(\bm{x}). \label{eq_bedingung_W2}
\end{align}
If $m \neq k$, then $W(\bm{x})$ must depend only on $x_k$ and $x_m$, while it can depend on all coordinates for $m = k$.
Then, a potential matrix of $\bm f$ is given by the $n \times n$ matrix-valued map $\mathbf{F} \in C^3(\mathbb{R}^n, \mathbb{R}^{n^2})$:
\begin{align}
\begin{split}
F_{ij}(\bm{x}) &\coloneqq \delta_{ik} \sum_{p=0}^{\lambda-1} \frac{(-1)^{p}}{\uphi} \partial_{x_j} \partial_{x_m}^{2\lambda - 2p - 2} \Delta_{\setminus m}^{p} W(\bm{x}). \\
\end{split} \label{eq_potential_Fkj_a}
\end{align}
To get the potential matrix of any linear combination of vector fields each satisfying the conditions above, calculate the matrices $\mathbf{F}$ term by term and sum them up.
\end{theorem}
\begin{proof}
Only the $k$-th row of $\mathbf{F}$ is non-zero, so gradient and rotation potentials yield:
\begin{align}
G(\bm x) &= F_{kk}(\bm x) = \sum_{p=0}^{\lambda-1} \frac{(-1)^p}{\uphi} \partial_{x_k} \partial_{x_m}^{2\lambda-2p-2} \Delta_{\setminus m}^p W(\bm x), \\
R_{ij}(\bm x) &= \begin{cases}
  + F_{kj}(\bm x) = \sum_{p=0}^{\lambda-1} \frac{(-1)^p}{\uphi} \partial_{x_j} \partial_{x_m}^{2\lambda-2p-2} \Delta_{\setminus m}^p W(\bm x) & \text{ for } i = k, j \neq k, \\
 - F_{ki}(\bm x) = \sum_{p=0}^{\lambda-1} \frac{(-1)^p}{\uphi} \partial_{x_i} \partial_{x_m}^{2\lambda-2p-2} \Delta_{\setminus m}^p W(\bm x) & \text{ for } i \neq k, j = k, \\
  0 & \text{ otherwise.}
  \end{cases}
\end{align}
Apart from the $k$-th row and column, the matrix $\mathbf{R}$ is zero.

Without loss of generality, we assume $k = 1$ in the following.
Using Eqs.~(\ref{eq_def_g}--\ref{eq_def_r}), the components of the gradient field $\bm g$ and the rotation field $\bm r$ are:
\begin{align}
g_1(\bm x) &= \sum_{p=0}^{\lambda-1} \frac{(-1)^p}{\uphi} \partial_{x_1}^2 \partial_{x_m}^{2\lambda-2p-2} \Delta_{\setminus m}^p W(\bm x), \label{eq_g_W} \\
r_1(\bm x) &= \sum_{j \neq 1} \sum_{p=0}^{\lambda-1} \frac{(-1)^p}{\uphi} \partial_{x_j}^2 \partial_{x_m}^{2\lambda-2p-2} \Delta_{\setminus m}^p W(\bm x), \label{eq_r_W} \\
g_{j \neq 1}(\bm x) &= + \sum_{p=0}^{\lambda-1} \frac{(-1)^p}{\uphi} \partial_{x_1} \partial_{x_j} \partial_{x_m}^{2\lambda-2p-2} \Delta_{\setminus m}^p W(\bm x), \\
r_{j \neq 1}(\bm x) &= - \sum_{p=0}^{\lambda-1} \frac{(-1)^p}{\uphi} \partial_{x_1} \partial_{x_j} \partial_{x_m}^{2\lambda-2p-2} \Delta_{\setminus m}^p W(\bm x).
\end{align}
Obviously, $g_{j \neq 1}(\bm x) + r_{j \neq 1}(\bm x) = 0$, which is equal to $f_{j \neq 1}(\bm x)$. To show that $g_1(\bm x) + r_1(\bm x) = f_1(\bm x)$, we have to distinguish the cases $m = 1$ and $m \neq 1$.

For $m = 1$, Eqs.~(\ref{eq_g_W}--\ref{eq_r_W}) yield:
\begin{align}
g_1(\bm x) &= \sum_{p=0}^{\lambda-1} \frac{(-1)^p}{\uphi} \partial_{x_1}^{2\lambda-2p} \Delta_{\setminus 1}^p W(\bm x), \label{eq_g1_m1} \\
r_1(\bm x) &= \sum_{j \neq 1} \sum_{p=0}^{\lambda-1} \frac{(-1)^p}{\uphi} \partial_{x_j}^2 \partial_{x_1}^{2\lambda-2p-2} \Delta_{\setminus 1}^p W(\bm x) = - \sum_{p=1}^\lambda \frac{(-1)^p}{\uphi} \partial_{x_1}^{2\lambda-2p} \Delta_{\setminus 1}^p W(\bm x). \label{eq_r1_m1}
\end{align}
In Eq.~\eqref{eq_r1_m1}, we replaced $\sum_{j \neq 1} \partial_{x_j}^2$ with $\Delta_{\setminus 1}$ and then shifted the remaining sum by replacing each $p$ with $p-1$.
Calculating $g_1(\bm x) + r_1(\bm x)$, the terms with equal index $p$ cancel each other, and the term with $p=0$ of $g_1$ and the term with $p = \lambda$ of $r_1$ remain. Using the conditions in Eqs.~(\ref{eq_bedingung_W1}--\ref{eq_bedingung_W2}) and $\Delta_{\setminus 1}^0 W = W$, it holds:
\begin{align}
 g_1(\bm x) + r_1(\bm x) = \frac{1}{\uphi} \partial_{x_1}^{2\lambda} W(\bm x) - \frac{(-1)^\lambda}{\uphi} \Delta_{\setminus 1}^\lambda W(\bm x) = \frac{1}{\uphi} f_1(\bm x) - \frac{1-\uphi}{\uphi} f_1(\bm x) = f_1(\bm x),
\end{align}
which proves that $\bm g(\bm x) + \bm r(\bm x) = \bm f(\bm x)$ for $m = 1$.

\todone{Reviewer: -p.12 l.15: "contributes only an element if" -> "contributes only if" \par Answer: Thanks.}
For $m \neq 1$, $W(\bm x)$ depends only on $x_1$ and $x_m$ by assumption, so $\Delta_{\setminus m}^p W(\bm x) = \partial_{x_1}^{2p} W(\bm x)$, and the sum over $j \neq 1$ in Eq.~\eqref{eq_r_W} contributes only if $j = m$. Therefore, Eqs.~(\ref{eq_g_W}--\ref{eq_r_W}) yield:
\begin{align}
g_1(\bm x) &= \sum_{p=0}^{\lambda-1} \frac{(-1)^p}{\uphi} \partial_{x_1}^{2p+2} \partial_{x_m}^{2\lambda-2p-2} W(\bm x) = - \sum_{p=1}^\lambda \frac{(-1)^p}{\uphi} \partial_{x_1}^{2p} \partial_{x_m}^{2\lambda-2p} W(\bm x), \label{eq_g_Wm} \\
r_1(\bm x) &= \sum_{p=0}^{\lambda-1} \frac{(-1)^p}{\uphi} \partial_{x_1}^{2p} \partial_{x_m}^{2\lambda-2p} W(\bm x),
\end{align}
shifting the sum in Eq.~\eqref{eq_g_Wm} by replacing each $p$ with $p-1$. Calculating $g_1(\bm x) + r_1(\bm x)$, the terms with equal index $p$ cancel each other, and the terms with $p=\lambda$ of $g_1$ and $p=0$ of $r_1$ remain. Using the conditions in Eqs.~(\ref{eq_bedingung_W1}--\ref{eq_bedingung_W2}) and $\partial_{x_1}^0 W = W$, it holds:
\begin{align}
g_1(\bm x) + r_1(\bm x) = \frac{1}{\uphi} \partial_{x_m}^{2\lambda} W(\bm x) - \frac{(-1)^\lambda}{\uphi} \Delta_{\setminus m}^\lambda W(\bm x) = \frac{1}{\uphi} f_1(\bm x) - \frac{1-\uphi}{\uphi} f_1(\bm x) = f_1(\bm x),
\end{align}
which proves that $\bm g(\bm x) + \bm r(\bm x) = \bm f(\bm x)$ for $m \neq 1$.

The statement on linear combinations follows immediately from the linearity of the Helmholtz decomposition (Lemma \ref{lemma_linear}), which concludes the proof of Theorem~\ref{theorem_hd_both}.
\end{proof}

The following three corollaries show how the conditions of Theorem~\ref{theorem_hd_both} can be satisfied for vector fields that are non-zero only in the first component.
Corollaries \ref{corol_green_1} and \ref{corol_green_2} deal with vector components that are partly described by a monomial, whereas Corollary \ref{corol_green_3} can be used for combinations of trigonometric and exponential functions.

\begin{corol} \label{corol_green_1}
Let $\bm{f}$ be a $C^1$ vector field of the form $\bm f = [q(\bm x), 0, \dots, 0]$ with $q$ being any function satisfying $\Delta_{\setminus 1}^\lambda q(\bm x) = 0$ with $\lambda \in \mathbb{N}_0$.
Then, a potential matrix of $\bm f$ can be derived using Theorem~\ref{theorem_hd_both} with $x_1$ as the $W$-coordinate, $\uphi = 1$ and $W(\bm x) = \mathcal{A}_{x_1}^{2 \lambda} f_1(\bm x)$.
\end{corol}
\begin{proof}
The conditions of Theorem~\ref{theorem_hd_both} are satisfied because $\partial_{x_1}^{2 \lambda} W(\bm x) = f_1(\bm x)$ and $(-1)^\lambda \Delta_{\setminus 1}^\lambda W(\bm x) = 0 = (1 - \uphi) f_1(\bm x)$.
\end{proof}
For example, if $\bm f = [q(x_1), 0, \dots, 0]$ depends only on $x_1$, Corollary~\ref{corol_green_1} can be used with $\lambda = 1$ to get: $F_{11}(\bm{x}) = \mathcal{A}_{x_1} q(x_1)$ and $F_{ij} = 0$ otherwise; $G(\bm{x}) = \mathcal{A}_{x_1} q(x_1)$; $R_{ij}(\bm{x}) = 0\ \forall\ i, j$; $\bm{g}(\bm{x}) = \bm{f}(\bm{x})$; $\bm{r}(\bm{x}) = 0$, implying that $\bm f$ is a gradient field.

\begin{corol} \label{corol_green_2}
Let $\bm{f}$ be a $C^1$ vector field of the form $\bm f = [x_1^b \cdot q(x_m), 0, \dots, 0]$ with $b \in \mathbb{N}_0$ and $q$ being any function dependent only on one coordinate $x_m \neq x_1$.
Then, a potential matrix of $\bm f$ can be derived using Theorem~\ref{theorem_hd_both} with $x_m$ as the $W$-coordinate, $\uphi = 1$, $\lambda = \lceil \tfrac{b + 1}{2} \rceil$ and $W(\bm x) = \mathcal{A}_{x_m}^{2 \lambda} f_1(\bm x)$.
\end{corol}
\begin{proof}
The conditions of Theorem~\ref{theorem_hd_both} are satisfied because $\partial_{x_m}^{2 \lambda} W(\bm x) = f_1(\bm x)$ and $(-1)^\lambda \Delta_{\setminus m}^\lambda W(\bm x) = (-1)^\lambda \partial_{x_1}^{2\lambda} W(\bm x) = 0 = (1 - \uphi) f_1(\bm x)$.
\end{proof}

If $\bm f = [q(x_m), 0, \dots, 0]$ and the field depends only on one `foreign' coordinate $x_m$, Corollary~\ref{corol_green_2} can be simplified with $b= 0$, $\lambda = 1$ to get: $F_{1m}(\bm{x}) = \mathcal{A}_{x_m} q(x_m)$ and $F_{jk}(\bm{x}) = 0$ otherwise; $G(\bm{x}) = 0$; $R_{1m}(\bm{x}) = -R_{m1}(\bm{x}) = \mathcal{A}_{x_m} q(x_m)$ and $R_{jk}(\bm{x}) = 0$ otherwise; $\bm{g}(\bm{x}) = 0$; $\bm{r}(\bm{x}) = \bm{f}(\bm{x})$, implying that $\bm f$ is a rotation field.

\begin{corol} \label{corol_green_3}
Let $\bm{f}$ be a $C^1$ vector field of the form $\bm f = [q(\bm x), 0, \dots, 0]$ with $\partial_{x_1}^2 q(\bm x) = v_1 q(\bm x)$ and $\Delta_{\setminus 1} q(\bm x) = v_2 q(\bm x)$ with $v_1, v_2$ independent of $\bm x$ and $v_1 \neq 0$.
Then, a potential matrix of $\bm f$ can be derived using Theorem~\ref{theorem_hd_both} with $x_1$ as the $W$-coordinate, $\uphi = 1 + v_2/v_1$, $\lambda = 1$ and $W(\bm x) = q(\bm x)/v_1$.
\end{corol}
\begin{proof}
The conditions of Theorem~\ref{theorem_hd_both} are satisfied because $\partial_{x_1}^2 W(\bm x) = \partial_{x_1}^2 (q(\bm x)/v_1) = (v_1 q(\bm x))/v_1 = f_1(\bm x)$ and $(-1)^\lambda \Delta_{\setminus 1} W(\bm x) = - v_2 q(\bm x) / v_1 = (1 - u) f_1(\bm x)$.
\end{proof}

\begin{figure}[tb]
\centering
\begin{tikzpicture}[node distance=2cm]

\node[align = center] (dec_ismonomxi) [decision] {Is $S \propto q(x_k) \cdot \bm x_{i \neq k}^{\beta}$ \\ with multi-index $\beta \in \mathbb{N}_0^{n-1}$?};

\newcommand{\ofx}{} 

\node[align = center] (dec_integratem) [decision, below of=dec_ismonomxi, left of = dec_ismonomxi, xshift=-1cm, yshift=-0.5cm] {Is $S\ofx \propto x_k^b x_i^{\beta_i}$ for \\ one $i \neq k$, $\beta_i > b$ \\ and $b, \beta_i \in \mathbb{N}_0$?};
\node[align = center] (dec_monom_xk) [decision, below of=dec_ismonomxi, xshift=2cm, yshift=-0.5cm] {Is $S\ofx \propto x_k^b \cdot q(x_i)$ \\ for one $i \neq k$ \\ and $b \in \mathbb{N}_0$?};

\draw [arrow] (dec_ismonomxi) -- node[anchor=east] {yes} (dec_integratem);
\draw [arrow] (dec_ismonomxi) -- node[anchor=west] {no} (dec_monom_xk);

\node[align = center] (dec_cossinexp) [decision, below of=dec_ismonomxi, right of = dec_ismonomxi, xshift=4.5cm, yshift=-0.5cm] {Is $\partial_{x_k}^2 S\ofx = v_1 S\ofx \neq 0$ \\ and $\Delta_{\setminus k} S\ofx = v_2 S\ofx$ \\ with $v_1, v_2$ constants?};
\draw [arrow] (dec_monom_xk) -- node[anchor=south] {no} (dec_cossinexp);

\node[align = left] (proc_integratek) [process, below of =  dec_integratem, yshift=-1.5cm, xshift=-1cm] {$\begin{aligned}m &= k, \\ \uphi &= 1,\\ \lambda &= \lceil \tfrac{|\beta| + 1}{2} \rceil, \\ W\ofx &= \mathcal{A}_{x_k}^{2 \lambda} S\ofx \end{aligned}$ };
\node[align = left] (proc_integratem) [process, below of = dec_monom_xk, yshift=-1.5cm, xshift=-2cm] {$\begin{aligned}m &= i,\\ \uphi &= 1,\\ \lambda &= \lceil \tfrac{b + 1}{2} \rceil,\\ W\ofx &= \mathcal{A}_{x_i}^{2 \lambda} S\ofx \end{aligned}$ };
\draw [arrow] (dec_integratem) -- node[anchor=west] {yes} (proc_integratem);
\draw [arrow] (dec_integratem) -- node[anchor=east] {no} (proc_integratek);
\draw [arrow] (dec_monom_xk) -- node[anchor=east] {yes} (proc_integratem);

\node[align = left] (proc_cossinexp) [process, below of = dec_cossinexp, yshift=-1.5cm, xshift=-2.5cm] {$\begin{aligned} m &= k,\\ \uphi &= 1 + v_2/v_1,\\ \lambda &= 1,\\ W\ofx &= S\ofx/v_1 \end{aligned}$ };
\draw [arrow] (dec_cossinexp) -- node[anchor=east] {yes} (proc_cossinexp);

\node[align = center] (solution) [startstop, below of = proc_integratem, yshift=-1cm] { $F_{ij}\ofx = \delta_{ik} \sum_{p=0}^{\lambda-1} \frac{(-1)^{p}}{\uphi} \partial_{x_j} \partial_{x_m}^{2\lambda - 2p - 2} \Delta_{\setminus m}^{p} W\ofx$. };

\node[align = left] (proc_noidea) [startstop, right of = solution, xshift=4.5cm] {No solution known.};

\draw [arrow] (dec_cossinexp) -- node[anchor=west] {no} (proc_noidea);
\draw [arrow] (proc_integratem) -- (solution);
\draw [arrow] (proc_integratek) -- (solution);
\draw [arrow] (proc_cossinexp) -- (solution);

\foreach \anc/\n in {north west/Q1}{\node[anchor=\anc, xshift=-0.6cm, yshift=+0.3cm] at (dec_ismonomxi.\anc) {\n};}
\foreach \anc/\n in {north west/Q2}{\node[anchor=\anc, xshift=-0.6cm, yshift=+0.3cm] at (dec_integratem.\anc) {\n};}
\foreach \anc/\n in {north west/Q3}{\node[anchor=\anc, xshift=-0.6cm, yshift=+0.3cm] at (dec_monom_xk.\anc) {\n};}
\foreach \anc/\n in {north west/Q4}{\node[anchor=\anc, xshift=-0.6cm, yshift=+0.3cm] at (dec_cossinexp.\anc) {\n};}

\foreach \anc/\n in {north west/Corol. \ref{corol_green_1}}{\node[anchor=\anc, xshift=-0.1cm, yshift=+0.43cm] at (proc_integratek.\anc) {\n};}
\foreach \anc/\n in {north west/Corol. \ref{corol_green_2}}{\node[anchor=\anc, xshift=-0.1cm, yshift=+0.43cm] at (proc_integratem.\anc) {\n};}
\foreach \anc/\n in {north west/Corol. \ref{corol_green_3}}{\node[anchor=\anc, xshift=-0.1cm, yshift=+0.43cm] at (proc_cossinexp.\anc) {\n};}
\foreach \anc/\n in {north west/Theorem \ref{theorem_hd_both}}{\node[anchor=\anc, xshift=-0.1cm, yshift=+0.43cm] at (solution.\anc) {\n};}

\end{tikzpicture}
\caption{\label{fig_flowchart}We suggest the following strategy to derive a Helmholtz decomposition for a vector field $\bm f(\bm x)$: (1) deal with every vector component $f_k(\bm x)$ separately, (2) expand each component into a sum of expressions $S(\bm x)$, (3) use the flowchart to find the appropriate corollary for each expression $S(\bm x)$ separately, (4) sum all the resulting matrices $F_{ij}(\bm x)$ in line with Lemma~\ref{lemma_linear}.
In the diagram, $q$ means any function of its arguments.}
\end{figure}

Based on these three corollaries, Figure~\ref{fig_flowchart} suggests a strategy to find the Helmholtz decomposition for any vector field.
A \emph{Mathematica} worksheet implementing this strategy can be found in \citet{glotzl_2023_helmholtzmath} and at \url{https://oliver-richters.de/helmholtz}.

Note that vector fields may satisfy the conditions of more than one corollary and the potential matrix is not uniquely defined,  see \ref{app_monomial} in the Appendix.
The flowchart in Figure~\ref{fig_flowchart} is set up such that, for the monomials, the solution with the smallest number of terms is selected.
It contains all strategies known to the authors to find a closed-form solution.
If no solution is known, the function $W$ may be obtained by line integration as $W(\bm x) = \mathcal{A}_{x_m}^{2\lambda} f_k(\bm x)$, but choosing the integration constants wisely is necessary to satisfy Eq.~\eqref{eq_bedingung_W2}.
Let us now apply the strategy from Figure~\ref{fig_flowchart} to a linear vector field.

\begin{example}[Linear vector fields] \label{sec_corollary_linear_fields}
Let $\bm{f} \in C^1(\mathbb{R}^n, \mathbb{R}^n)$ be a linear vector field
\begin{align}
f(\bm{x}) &= \mathbf{L} \bm{x} \quad \Leftrightarrow \quad f_k(\bm{x}) = \sum_{j=1}^n L_{kj} x_j,
\end{align}
with an $n \times n$ matrix $\mathbf{L} = \llbracket L_{ij} \rrbracket$.
Each summand can be treated separately.
For a term $L_{kk} x_k$, Figure~\ref{fig_flowchart} suggests Corollary~\ref{corol_green_1}, whereas for non-diagonal terms $L_{kj} x_j$, Corollary~\ref{corol_green_2} is recommended.

The potentials and the vector fields constituting a Helmholtz decomposition are:
\begin{align}
F_{ij}(\bm{x}) &= \tfrac12 L_{ij} x_j^2,  \label{eq_corro_5a} \\
G(\bm{x}) &= \sum_{k=1}^n \tfrac12 L_{kk} x_k^2, \\
R_{ij}(\bm{x}) &= \tfrac12 L_{ij} x_j^2 - \tfrac12 L_{ji} x_i^2, \\
g_k(\bm{x}) &= L_{kk} x_k, \\
r_k(\bm{x}) &= \sum_{j\neq k} L_{kj} x_j. \label{eq_corro_5b}
\end{align}
\end{example}
The linear vector field $\bm{f}(\bm{x}) = \left( \begin{smallmatrix} +1 & +1 \\ -1 & +1 \end{smallmatrix} \right) \bm{x} = \left[ x_1 + x_2, x_2 - x_1 \right]$ can be decomposed into $\bm g(\bm{x}) = [+ x_1, + x_2]$ and $\bm r(\bm{x}) = [+ x_2, - x_1]$ using $\mathbf{F}(\bm{x}) =  \begin{pmatrix} \frac12 x_1^2 & \frac12 x_2^2 \\ - \frac12 x_1^2 & \frac12 x_2^2 \end{pmatrix}$, $G(\bm{x}) = \frac12 (x_1^2 + x_2^2)$, and $\mathbf{R}(\bm{x}) = \begin{pmatrix} 0 & \frac12 (x_1^2 + x_2^2) \\  - \frac12 (x_2^2 + x_1^2) & 0 \end{pmatrix}$.
Further examples are presented in the Appendix.

\section{Helmholtz decomposition for analytic vector fields}
\label{sec_hd_analytic}

The next step is to prove the existence of a Helmholtz decomposition for all analytic vector fields by showing that the potential matrix $\mathbf{F}$ of an analytic vector field $\bm{f}$ is analytic as well.
In Lemma~\ref{lemma_monomial}, we first establish some upper bounds for the polynomials appearing in the potential matrix for a vector field only containing one monomial in one of its coordinates.

As a notation, we use $\sum_\gamma c_{\gamma} \bm x^\gamma$ to describe a power series, with $\gamma$ a multi-index, $|c_{\gamma}|$ the sum of the absolute values of the coefficients, and $|c_{|\gamma| = D}|$ the sum of the absolute values of the coefficients of monomials with degree $D$.

\begin{lemmarev} \label{lemma_monomial}
Let a vector field $\bm{f} \in C^\infty(\mathbb{R}^n, \mathbb{R}^n)$ consisting of one monomial of degree $D$ be given by $\bm{f}(\bm{x}) = \Big[ x_1^b \cdot \bm x_{i \neq 1}^{\beta}, 0, \dots, 0 \Big]$,
with $b \in \mathbb{N}_0$, $\beta \in \mathbb{N}_0^{n-1}$ a multi-index and $D = b+ |\beta|$ with $|\beta|$ the degree of $\bm x_{i \neq 1}^{\beta}$.
Then, each component of the potential matrix of $\bm f$ is a polynomial $F_{ij}(\bm x) = \sum_\gamma c_{ij; \gamma} \bm x^\gamma$ (with $\gamma \in \mathbb{N}_0^n$ a multi-index) consisting of monomials with degree $D+1$ only, and $\left| c_{ij; \gamma} \right| \leq (D+1) n^{D+1} \sqrt{\frac{2 (D+1)}{\pi}} 2^{D+2}$.

\begin{proof}
To derive a potential matrix, we note that the conditions of Theorem \ref{theorem_hd_both} are satisfied with $x_1$ as the $W$-coordinate,  $\uphi = 1$, $\lambda = \lceil (|\beta|+1)/2 \rceil$, and $W(\bm x) = \mathcal{A}_{x_1}^{2 \lambda} f_1(\bm x) = \frac{b!}{(b+2\lambda)!} x_1^{b+2\lambda} \bm x_{i \neq 1}^\beta$. Note that $\lambda \leq D + 1$.
Then, the potential matrix $\mathbf{F}$ has the following components:
\begin{align}
\label{eq_F_monomial}
F_{ij}(\bm{x}) &= \delta_{i1} \sum_{p=0}^{\lambda-1} (-1)^{p} \partial_{x_j} \partial_{x_1}^{2\lambda - 2p - 2} \Delta_{\setminus 1}^{p} W(\bm{x}).
\end{align}
$W$ is a monomial with degree $D + 2\lambda$. Each monomial in the sum of Eq.~\eqref{eq_F_monomial} has degree $D+1$, as it is calculated by applying $1 + (2\lambda - 2 p - 2) + 2 p = 2 \lambda - 1$ derivatives on $W$.
The total number of monomials $M$ in Eq.~\eqref{eq_F_monomial} is smaller than
\begin{align}
M &= \lambda \cdot n^\lambda \leq (D+1) \cdot n^{D+1}, \label{eq_monomial_M}
\end{align}
the first $\lambda$ covering the sum, the $n^\lambda$ covering the repeated application of the Laplacian, where each term is split up into at most $n$ summands, a maximum of $\lambda$ times.

For each term in the sum of Eq.~\eqref{eq_F_monomial}, differentiating leads to coefficients, that are for each $p$ bounded by:
\begin{align}
P &= (b + |\beta|) \cdot \frac{ b! }{(b + 2p + 2)!} \cdot \frac{|\beta|!}{(|\beta| - 2 p)!}. \label{eq_P_upperbound}
\end{align}
The first term is an upper limit for the pre-factor caused by the differentiation $\partial_{x_j}$, the second term the exact coefficient of $\partial_{x_1}^{2\lambda - 2p - 2} \frac{b!}{(b+2\lambda)!} x_1^{b+2\lambda}$, and the third term is an upper limit for differentiating $2 p$ times a monomial $\bm x_{i \neq 1}^\beta$ that has exponents not bigger than $|\beta|$.

We now establish an upper limit for $P$.
The product of the factorials in the denominators can be written as $\Gamma(b + 2 p + 3) \cdot \Gamma(|\beta| - 2 p + 1)$ and, 
by the log-convexity of the gamma function, it follows that:
\begin{align}
\Gamma(b + 2 p + 3) \cdot \Gamma(|\beta| - 2 p + 1) \geq \Gamma\left( \frac{b + |\beta| + 4}{2} \right)^2 \geq \Gamma\left(\frac{D+1}{2}\right)^2.
\end{align}
The product of the numerators has an upper limit as follows:
\begin{align}
b! \cdot |\beta|! &\leq (b + |\beta|)! = D! = \Gamma(D + 1).
\end{align}
Therefore, an upper bound for $P$ can be derived:
\begin{align}
P &\leq (D+1) \frac{\Gamma(D+1)}{\Gamma((D+1)/2)^2}.
\end{align}
If the degree of the monomial $D$ becomes large, the gamma function can be approximated using Stirling's formula that $\Gamma(D) \approx \sqrt{2 \pi D}\left(\frac{D}{e}\right)^D$:
\begin{align}
P &\leq 2 (D+1) \frac{ \sqrt{ 2 \pi (D+1) } \left(\frac{D+1}{e}\right)^{D+1}}{ \left(\sqrt{ 2 \pi (D+1) / 2} \left(\frac{D+1}{2e}\right)^{(D+1)/2}\right)^2} = \sqrt{\frac{2 (D+1)}{\pi}} 2^{D+2}. \label{eq_monomial_P}
\end{align}
The first factor $2$ is a safety margin, making sure that the inequality holds despite the fact that the Stirling formula slightly misestimates the factorial.

Taking Eqs.~\eqref{eq_monomial_M} and \eqref{eq_monomial_P} together, the absolute values of the coefficients of $F_{ij} = \sum_\gamma c_{ij; \gamma} \bm{x}^\gamma$ are bounded as follows:
\begin{align}
\left| c_{ij; \gamma} \right| \leq (D+1) n^{D+1} \sqrt{\frac{2 (D+1)}{\pi}} 2^{D+2}.
\end{align}
\end{proof}
\end{lemmarev}

\begin{theorem} \label{theorem_hd_analytic}
Let $\bm f$ be an analytic vector field with a radius of convergence $\rho$.
Then, a local Helmholtz decomposition of $\bm{f}$ exists, i.\,e., there exist a potential matrix $\mathbf{F}$, a gradient potential $G$, a rotation potential $\mathbf{R}$, a gradient field $\bm{g}$ and a rotation field $\bm{r}$ that are analytic with a strictly positive radius of convergence $\rho^H$. If $\rho$ is infinite, then $\rho^H$ is infinite as well.
\end{theorem}

\begin{proof}
We first prove the theorem for $\bm f$ that has the form $\left[ f_1(\bm x), 0, \dots, 0 \right]$.
The non-zero component $f_1$ is given locally around $\bm \xi$ by the convergent power series:
\begin{align}
f_1(\bm{x}) = \sum_\alpha a_{\alpha} (\bm x - \bm \xi)^{\alpha}, \label{eq_powerseries}
\end{align}
using the multi-index $\alpha \in \mathbb{N}_0^n$.
Without loss of generality, we prove the existence for $\bm \xi = 0$.
According to the Cauchy–Hadamard theorem \citep[p.~32]{shabat_introduction_1992}, the radius of convergence $\rho > 0$ satisfies:
\begin{align}
\limsup_{D \to\infty} \sqrt[D]{\left| a_{|\alpha| = D} \right|} = 1/\rho. \label{eq_radius_convergence}
\end{align}
According to Lemma~\ref{lemma_monomial}, each monomial in the power series of $f_1$ with degree $D$ has a potential matrix with components that are polynomials consisting of monomials of degree $D+1$.
The absolute values of the polynomials' coefficients sum at most to $(D+1) n^{D+1} \sqrt{\frac{2 (D+1)}{\pi}} 2^{D+2}$.
To derive the potential matrix $\mathbf{F}$ of $\bm f$, one has to sum over the potential matrices of all monomials, yielding a power series:
\begin{align}
F_{ij}(\bm x) &= \delta_{1i} \sum\limits_{\gamma} c_{ij;\gamma} \cdot \bm x^\gamma,
\end{align}
using the multi-index $\gamma \in \mathbb{N}_0^{n}$.
We will show that the radius of convergence of $\mathbf F$ is positive by relating the coefficients of the power series of $\bm f$ to those of $\mathbf F$.

For a given degree $|\gamma| = D+1$, an upper limit for $|c_{ij; |\gamma| = D+1}|$ is given by
\begin{align}
\left| c_{ij; |\gamma| = D+1} \right| \leq \sqrt{\frac{2 (D+1)}{\pi}} 2^{D+2} \cdot (D+1) n^{D+1} \cdot \left|a_{|\alpha| = D}\right|.
\end{align}
This way, we can establish a lower bound for the radius of convergence $\rho^H$ for the potential matrix $\mathbf{F}$ in relation to the radius of convergence $\rho$ of the original vector field $\bm{f}$ as given by Eq.~\eqref{eq_radius_convergence}:
\begin{align}
1/\rho^H_{ij} &= \limsup_{D \to\infty} \sqrt[D+1]{\left| c_{ij;|\gamma| = D+1} \right|} \\
 &\leq \limsup_{D\to\infty} \sqrt[D+1]{\sqrt{\frac{2 (D+1)}{\pi}} 2^{D+2} (D+1) n^{D+1} \left| a_{|\alpha| = D} \right|} \\
 &= 2 n / \rho. \label{eq_radius_convergence_new}
\end{align}
The last step takes into account that $\limsup_{D \to\infty} \sqrt[D+1]{\sqrt{\frac{2 (D+1)}{\pi}} 2 (D+1)} = 1$. 
As this formula is independent of the index $ij$, the analytic functions describing the matrix $\mathbf{F}$ have a lower limit for the radius of convergence $\rho^H$ of $\rho/2n$.
Therefore, also the matrix $\mathbf{F}$ and the vector fields $\bm{g}$ and $\bm{r}$ that are obtained as derivatives of $\mathbf{F}$ are analytic and their radius of convergence is strictly positive.
Vector fields where every coordinate is an `entire function' with an infinite radius of convergence $\rho$ have a Helmholtz decomposition with an infinite radius of convergence $\rho^H$ as well.

The other components $f_{i \neq 1}(\bm x)$ work analogously, and summing up the potential matrices $\mathbf{F}$ in line with Lemma~\ref{lemma_linear} yields the Helmholtz decomposition of $\bm f(\bm x)$.
\end{proof}

\section{Discussion and conclusions}
\label{sec_conclusion}

This paper introduces the \emph{rotation potential} $\mathbf{R} = \llbracket R_{ij} \rrbracket$ for an $n$-dimensional vector field $\bm{f}$ as a generalization of the three-dimensional \emph{vector potential} $\bm{A}$.
Together with the \emph{gradient potential} $G$, the $n \times n$ matrix-valued map $\mathbf{R}$ can be used to derive a Helmholtz decomposition with a rotation-free \emph{gradient field} $\bm{g}(\bm{x}) = \grad G(\bm{x})$ and a divergence-free \emph{rotation field} $\bm{r}(\bm{x}) = \ROT \mathbf{R}(\bm x)$ using a \emph{rotation operator} $\ROT$.
The antisymmetric matrix $\mathbf{R}$ is the sum of the rotation potentials for the $\binom{n}{2}$ rotations within the coordinate planes.

Theorem~\ref{theorem_hd_decay} derives the potentials for fields decaying sufficiently fast at infinity using convolution integrals.
Theorem~\ref{theorem_hd_both} provides closed-form solutions for many unbounded, smooth vector fields that satisfy certain sufficient (and probably necessary) conditions that allow using a technique based on line integrals.
Theorem~\ref{theorem_hd_analytic} proves that this method can be extended to all analytic vector fields.
A linear combination of vector fields that all satisfy the conditions of one of the theorems can be decomposed by deriving the potentials element-wise, and summing them up.
The procedure of Theorem~\ref{theorem_hd_both} is illustrated in the Appendix for multivariate polynomials, exponential functions, trigonometric functions, and three examples from complex system theory, the R\"{o}ssler and Lorenz attractors, and generalized Lotka--Volterra equations.

In any case, it has to be kept in mind that a Helmholtz decomposition is never unique:
adding a harmonic function $H \in C^2(\mathbb{R}^n, \mathbb{R})$ with $\Delta H(\bm{x}) = 0$ to the gradient potential $G$, and correcting $\bm{r}$ to keep $\bm{g} + \bm{r} = \bm{f}$ satisfied, yields the vector fields
\begin{align}
 \bm{g^\prime}(\bm{x}) &= \grad (G(\bm{x}) + H(\bm{x})), \\
 \bm{r^\prime}(\bm{x}) &= \bm{r}(\bm{x}) - \grad H(\bm{x}),
\end{align}
that are also a Helmholtz decomposition of $\bm{f}$.
By choosing this harmonic function $H$, additional boundary conditions can be satisfied, known as `gauge fixing' in physics.

Thanks to their versatility and the possibility of obtaining closed-form solutions for the Helmholtz decomposition, the theorems presented in this paper may prove helpful for problems of vector analysis, theoretical physics, and complex systems theory.


\section*{Acknowledgments}
Author names in alphabetical order.
EG thanks Walter Zulehner from Johannes Kepler University Linz. OR thanks Ulrike Feudel and Jan Freund from Carl von Ossietzky University of Oldenburg. Both authors thank the anonymous reviewers and Rebecca Klinkig for their valuable comments. All errors and omissions are our own.

\renewcommand{\bibfont}{\normalfont\small}
\addcontentsline{toc}{section}{References} 

\printbibliography

\clearpage

\begin{appendices}

\renewcommand{\thesubsection}{Example \arabic{subsection}}

\section{Examples}

Each example can be derived by following the strategy in Figure~\ref{fig_flowchart}. A \emph{Mathematica} worksheet can be found at \url{https://oliver-richters.de/helmholtz} and in \citet{glotzl_2023_helmholtzmath}.

\subsection{Multivariate monomial}
\label{app_monomial}

Consider a vector field $\bm{f}$ with a multivariate monomial given by
\begin{align}
\bm{f}(\bm{x}) = \Big[x_1^2 x_2^{20}, 0\Big].
\end{align}
With $x_2$ as the $W$-coordinate, $\lambda = 2$, $W(\bm x) = \frac{20!}{24!} x_1^2 x_2^{24}$ and $\uphi = 1$, the conditions of Theorem~\ref{theorem_hd_both} are satisfied, as $\Delta_{\setminus 2}^{2\lambda} W(\bm x) = \partial_{x_1}^4 W(\bm x) = 0$. The Helmholtz decomposition is given by:
\begin{align}
F_{ij}(\bm{x}) &= \begin{pmatrix} \frac{2 \cdot 20!}{22!} x_1 x_2^{22} & \frac{1}{21} x_1^2 x_2^{21} - \frac{2 \cdot 20!}{23!} x_2^{23}  \\ 0 & 0 \end{pmatrix}, \\
G(\bm{x}) &= \frac{2 \cdot 20!}{22!} x_1 x_2^{22}, \\
R_{ij}(\bm{x}) &= \begin{pmatrix} 0 & + \frac{1}{21} x_1^2 x_2^{21} - \frac{2 \cdot 20!}{23!} x_2^{23} \\ - \frac{1}{21} x_1^2 x_2^{21} + \frac{2 \cdot 20!}{23!} x_2^{23} & 0 \end{pmatrix}, \\
\bm{g}(\bm{x}) &= \left[\hphantom{x_1^2 x_2^{20}} + \frac{2 \cdot 20!}{22!} x_2^{22}, + \frac{2}{21} x_1 x_2^{21} \right], \\
\bm{r}(\bm{x}) &= \left[ x_1^2 x_2^{20} - \frac{2 \cdot 20!}{22!} x_2^{22}, - \frac{2}{21} x_1 x_2^{21} \right].
\end{align}
One could also choose $x_1$ as the $W$-coordinate, $\lambda = 21$, $W(\bm x) = \frac{2}{44!} x_1^{44} x_2^{20}$ and $\uphi = 1$, but the resulting potential matrix $\mathbf{F}$ has 11 summands in each component, making it less tractable. For a vector field such as $\bm f(\bm x) = \left[x_1^2 x_2^{20} x_3^2, 0, 0 \right]$ on the other hand, there is no alternative to integrating $x_1$, because one cannot pick $x_2$ as the $W$-coordinate as $f_1(\bm x)$ depends not only on $x_1$ and $x_2$, which is a necessary condition of Theorem~\ref{theorem_hd_both} for $k \neq m$. The lengthy results can be found at \url{https://oliver-richters.de/helmholtz} and in \citet{glotzl_2023_helmholtzmath}.

\subsection{Cosine and Exponential function}
Consider a vector field $\bm{f} \in C(\mathbb{R}^2, \mathbb{R}^2)$ given by:
\begin{align}
\bm{f}(\bm{x}) = \Big[f_1(\bm x), 0\Big] = \Big[\cos(w x_1) e^{a x_2}, 0\Big],
\end{align}
with $a,w$ non-zero real numbers with $|a| \neq |w|$. As $\partial_{x_1}^2 f_1(\bm x) = -w^2 f_1(\bm x)$ and $\Delta_{\setminus 1} f_1(\bm x) = a^2 f_1(\bm x)$, the conditions of Corollary~\ref{corol_green_3} are satisfied, so Theorem~\ref{theorem_hd_both} can be used with $x_1$ as the $W$-coordinate, $W(\bm x) = - \frac{1}{w^2} \cos(w x_1) e^{a x_2}$, $u = 1-w^2/a^2$ and $\lambda = 1$:
\begin{align}
(-1)^{\lambda} \Delta_{\setminus 1}^{\lambda} W(\bm x) = \frac{a^2}{w^2} \cos(w x_1) e^{a x_2} = (1 - \uphi) f_1(\bm x). \label{eq_app_cosexp_condition}
\end{align}
Note that $W(\bm x)$ is here \emph{not} defined as $\mathcal{A}_{x_m}^2 f_1(\bm x) = - \frac{1}{w^2} (\cos(w x_1) - 1) e^{a x_2}$, but the integration constants were chosen such that Eq.~\eqref{eq_app_cosexp_condition} can be satisfied.
The Helmholtz decomposition is given by:
\begin{align}
\mathbf{F}(\bm x) &= \begin{pmatrix} \frac{w}{w^2-a^2} \sin(w x_1) e^{a x_2} & -\frac{a}{w^2-a^2} \cos(w x_1) e^{a x_2} \\ 0 & 0 \end{pmatrix}, \\
G(\bm x) &=  \frac{w}{w^2-a^2} \sin(w x_1) e^{a x_2}, \\
\mathbf{R}(\bm x) &= \begin{pmatrix} 0 & - \frac{a}{w^2 - a^2} \cos(w x_1) e^{a x_2} \\ + \frac{a}{w^2 - a^2} \cos(w x_1) e^{a x_2} & 0 \end{pmatrix}, \\
\bm{g}(\bm{x}) &= \left[ \frac{w^2}{w^2-a^2} \cos(w x_1) e^{a x_2}, \ \frac{+a w}{w^2-a^2} \sin(w x_1) e^{a x_2} \right], \\
\bm{r}(\bm{x}) &= \left[ \frac{-a^2}{w^2-a^2} \cos(w x_1) e^{a x_2}, \ \frac{-a w}{w^2-a^2} \sin(w x_1) e^{a x_2} \right].
\end{align}

\subsection{R\"{o}ssler attractor}

The R\"{o}ssler attractor, a classic example from complex system theory \citep{rossler_equation_1976, peitgen_strange_1992}, is given by:
\begin{align}
\bm{\dot x}= \bm{f}(\bm{x}) = \big[- x_2 - x_3, x_1 + a x_2, b - c x_3 + x_1 \big].
\end{align}
Treating each vector component individually, expanding each component into a sum and deriving the potentials for each summand using Figure~\ref{fig_flowchart} yields the Helmholtz decomposition:
\begin{align}
\mathbf{F}(\bm{x}) &= \begin{pmatrix} 0 & -\tfrac12 x_2^2 & -\tfrac12 x_3^2 \\[0.5em] \tfrac12 x_1^2 & \tfrac{a}{2} x_2^2 & 0 \\[0.5em] \tfrac16 x_3^3 & 0 & b x_3 - \tfrac{c}{2} x_3^2  + \tfrac12 x_1 x_3^2 \end{pmatrix}, \\
G(\bm{x}) &= \tfrac{a}{2} x_2^2 + b x_3 - \tfrac{c}{2} x_3^2 + \tfrac{1}{2} x_1 x_3^2, \\
\mathbf{R}(\bm{x}) &= \begin{pmatrix} 0 & - \tfrac12 x_1^2 -\tfrac12 x_2^2  & -\tfrac12 x_3^2 - \tfrac16 x_3^3 \\[0.5em] + \tfrac12 x_1^2 + \tfrac12 x_2^2  & 0 & 0 \\[0.5em] + \tfrac12 x_3^2 + \tfrac16 x_3^3 & 0 & 0 \end{pmatrix}, \\
\bm{g}(\bm{x}) &= \big[\hphantom{ -x_2 - x_3\ } + \tfrac12 x_3^2 , + a x_2, b - c x_3 + x_1 x_3 \big], \\
\bm{r}(\bm{x}) &= \big[-x_2 - x_3 - \tfrac12 x_3^2, + x_1 \hphantom{a}, \hphantom{b- x_3\,} 0 \hphantom{ + x_1 x_3}  \big].
\end{align}

\subsection{Lorenz attractor}

The Lorenz system, a simplified mathematical model for atmospheric convection with a strange attractor \citep{peitgen_strange_1992, lorenz_deterministic_1963}, is given by:
\begin{align}
\bm{\dot x} = \bm{f}(\bm{x}) = \big[a (x_2-x_1), x_1 (b-x_3)-x_2, x_1 x_2-c x_3 \big].
\end{align}
Treating each vector component individually, expanding each component into a sum and deriving the potentials for each summand using Figure~\ref{fig_flowchart} yields the Helmholtz decomposition:
\begin{align}
\mathbf{F}(\bm{x}) &= \begin{pmatrix} -\tfrac{a}{2} x_1^2  &   \tfrac{a}{2} x_2^2   &  0 \\[0.5em]  \tfrac{b}{2} x_1^2 - \tfrac12 x_2^2 x_3 & - \tfrac12 x_2^2  -x_1 x_2 x_3   &    - \tfrac12 x_2^2 x_1   \\[0.5em]  \tfrac12 x_2 x_3^2  & \tfrac12 x_1 x_3^2 & - \tfrac{c}{2} x_3^2 + x_1 x_2 x_3 \end{pmatrix}, \\
G(\bm{x}) &= -\tfrac{a}{2} x_1^2  - \tfrac12 x_2^2 - \tfrac{c}{2} x_3^2, \\
\mathbf{R}(\bm{x}) &= \begin{pmatrix}
0                                                                &    \tfrac{a}{2} x_2^2 - \tfrac{b}{2} x_1^2  + \tfrac12 x_2^2 x_3  &  - \tfrac12 x_2 x_3^2  \\[0.5em]  
 \tfrac{b}{2} x_1^2 - \tfrac{a}{2} x_2^2 - \tfrac12 x_2^2 x_3   &   0                                                               &    - \tfrac12 x_1 x_2^2  - \tfrac12 x_1 x_3^2  \\[0.5em]
+ \tfrac12 x_2 x_3^2                                             &  + \tfrac12 x_1 x_2^2 + \tfrac12 x_1 x_3^2                        &   0 
   \end{pmatrix}.  \\ 
\bm{g}(\bm{x}) &= \big[-a x_1, \hspace{2.9em}-x_2, -c x_3 \big], \\
\bm{r}(\bm{x}) &= \big[+ a x_2, b x_1 - x_1 x_3, \hspace{0.2em}x_1 x_2 \big].
\end{align}
The Lorenz system contains a square gradient potential, pushing the dynamics into the direction of the origin.
This is responsible for the stable fixed point at the origin for some parameters.
If the rotation field $\bm{r}$ becomes `strong' enough, it can push the dynamics away from this fixed point, creating a strange attractor.

\subsection{Competitive Lotka--Volterra equations with n species}

The $n$-species competitive Lotka--Volterra system, an important model in population dynamics \citep{wangersky_lotka-volterra_1978, montes_de_oca_balancing_1995}, is given by:
\begin{align}
\bm{\dot x} = \bm{f}(\bm{x}) &= x_i \left[ \rho_i - \sum_{j=1}^n \alpha_{ij} x_j ; \,_{1 \leq i \leq n} \right]. 
\end{align}
Treating each vector component individually, expanding each component into a sum and deriving the potentials for each summand using Figure~\ref{fig_flowchart}, yields the Helmholtz decomposition:
\begin{align}
F_{ik}(\bm x) &= \begin{cases}
  \tfrac12 \rho_k x_k^2 - \tfrac13 \alpha_{kk} x_k^3 - \tfrac12 x_k^2 \sum_{j \neq k} \alpha_{kj} x_j & \text{ for } i = k, \\
  -\tfrac16 x_i^3 \alpha_{ik} & \text{ otherwise},
\end{cases} \\
G(\bm{x}) &= \sum_{k=1}^n \left[ \tfrac12 \rho_k x_k^2 - \tfrac13 \alpha_{kk} x_k^3 - \tfrac12 x_k^2 \sum_{j \neq k} \alpha_{kj} x_j \right], \\
R_{ik}(\bm{x}) &= -\tfrac16 \alpha_{ik} x_i^3  + \tfrac16  \alpha_{ki} x_k^3, \\
\bm{g}(\bm{x}) &= \left[ \rho_i x_i - x_i \sum_{j=1}^n \alpha_{ij} x_j - \sum_{k \neq i} \tfrac12  \alpha_{ki} x_k^2; \,_{1 \leq i \leq n} \right], \\
\bm{r}(\bm{x}) &= \left[ \sum_{k\neq i} \tfrac12  \alpha_{ki} x_k^2; \,_{1 \leq i \leq n} \right].
\end{align}
The periodic oscillations for which these models are known arise from the interaction terms in the rotation potential.

\end{appendices}

\end{document}